\documentclass{svproc}
\usepackage{url}
\usepackage{graphicx}        
\usepackage{amsfonts}
\usepackage{cite}
\usepackage{amsmath}
\usepackage{comment}
\usepackage{xcolor}

\usepackage{lscape}

\usepackage{multirow}

\usepackage{array}
\newcolumntype{P}[1]{>{\centering\arraybackslash}p{#1}}

\begin{document}
\mainmatter              
\title{Analyzing Community-aware Centrality Measures Using The Linear Threshold Model}
\titlerunning{Analyzing Community-aware Centrality Measures}  
%
\author{Stephany Rajeh\inst{1} \and  Ali Yassin\inst{2} \and
	Ali Jaber\inst{2} \and
	Hocine Cherifi \inst{1}}
\authorrunning{S. Rajeh et al.} 
%
%

\institute{Laboratoire d’Informatique de Bourgogne - University of Burgundy, Dijon, France \\
\and
Computer Science Department - Lebanese University, Beirut, Lebanon\\}

\maketitle    

\begin{abstract}
Targeting influential nodes in complex networks allows fastening or hindering rumors, epidemics, and electric blackouts. Since communities are prevalent in real-world networks, community-aware centrality measures exploit this information to target influential nodes. Researches show that they compare favorably with classical measures that are agnostic about the community structure. Although the diffusion process is of prime importance, previous studies consider mainly the famous Susceptible-Infected-Recovered (SIR) epidemic propagation model. This work investigates the consistency of previous analyses using the popular Linear Threshold (LT) propagation model, which characterizes many spreading processes in our real life. We perform a comparative analysis of seven influential community-aware centrality measures on thirteen real-world networks. Overall, results show that Community-based Mediator, Comm Centrality, and Modularity Vitality outperform the other measures. Moreover, Community-based Mediator is more effective on a tight budget (i.e., a small fraction of initially activated nodes), while Comm Centrality and Modularity Vitality perform better with a medium to a high fraction of initially activated nodes.
 
\end{abstract}

\section{Introduction}
Complex systems are pervasive in biological, technological, and social sciences. Networks ideally represent these systems \cite{ doi:10.1080/00018732.2011.572452,rital2002combinatorial, rital2005weighted}. Identifying influential nodes has broad applications, from enhancing the diffusion process of information and marketing campaigns to controlling the epidemic spreading through immunization. One uses centrality measures to tackle this challenge \cite{das2018study}. The classical taxonomy considers two types of centrality measures. Local centrality measures, such as the degree centrality, exploit the neighborhood of a node. They are computationally efficient, yet they aren't always effective in identifying influential nodes. Global centrality measures, such as betweenness centrality, consider the overall network topology to compute the centrality of a node. Exploiting more information, they are generally more effective than local ones. However, they are computationally demanding. One can also combine both to counterbalance their pros and cons \cite{ibnoulouafi2018m}. 

Networks of interest in social, infrastructural, metabolic, and regulatory networks are divided naturally into communities \cite{girvan2002community}. These communities significantly affect the structure and the dynamics of the network. Recently developed centrality measures exploit this information to identify influential nodes \cite{ghalmane2019immunization, guimera2005functional, tulu2018identifying, gupta2016centrality, modvitality, zhao2015community, luo2016identifying}. Unlike classical centralities, community-aware centrality measures differentiate intra-community links (links between nodes in the same community) and inter-community links (links between nodes belonging to different communities) \cite{rajeh2021characterizing}. They offer different ways to mix intra-community and inter-community links. For instance, Community Hub-Bridge \cite{ghalmane2019immunization} combines the number of inter-community links weighted by the number of neighboring communities reached within one hop and the intra-community links weighted by the size of the node's community. Community-based Mediator \cite{tulu2018identifying} targets nodes based on how mixed the node's intra-community and inter-community links are.

Many diffusion models exist, each with different dynamics. So far, previous studies investigate community-aware centrality measures using the Susceptible-Infected-Recovered (SIR) epidemic propagation model or look at the network robustness \cite{luo2016identifying, modvitality, zhao2015community, gupta2016centrality, tulu2018identifying, ghalmane2019immunization}. Starting with a set of initial spreaders defined by a particular centrality measure, they evaluate its spreading efficiency with the maximum epidemic outbreak size obtained in a SIR simulation process. Another alternative is to evaluate the effectiveness of the centrality measure by the ability of the set of initial spreaders to break the network into multiple components. Consequently, the largest connected component is an upper bound on the outbreak size independent of the diffusion process. In practice, the influence of a node depends on the network topology and the underlying dynamics of propagation. Therefore, this paper aims to evaluate the performance of the community-aware centrality measures using one of the most popular diffusion processes: The Linear Threshold (LT) model. Indeed, originally developed to analyze collective behavior, LT and its variations are ubiquitous in many real-world situations \cite{granovetter1978threshold}. An extensive investigation of the most popular community-aware centrality measures is performed on thirteen real-world networks from different domains using the Linear Threshold model.

The paper is organized as follows. Section 2 introduces the community-aware centrality measures under test. Section 3 presents the Linear Threshold model. Section 4 reports basic information about the data and the evaluation measures. Section 5 provides the experimental results about the influence of the Linear Threshold model parameters on the diffusion process. Section 6 compares the centrality measures effectiveness across networks. Section 7 discusses the main findings. Finally, section 8 concludes the paper.




\section{Community-aware Centrality Measures}
\label{sec:Def}
In this section, the most commonly used community-aware centrality measures are defined. Each graph $G$ of $N$ nodes is divided into $C=\{c_1, c_2, ... , c_q, ..., c_{N_c}\}$ communities, where $N_c$ is the total number of communities, $c_q$ is the $q$-th community, and $n_{c_q}$ is the number of nodes in community $c_q$. In a given community, the total links of each node $i$ is represented by $k_{i,c_q}$. Furthermore, each node $i$ possesses $k_i^{intra}$ (intra-community links) and $k_i^{inter}$ (inter-community links). Consequently, the node's total degree is $k_i$ = $k_i^{intra}$ + $k_i^{inter}$.

\textbf{Comm Centrality} \cite{gupta2016centrality}: targets hubs in communities through the intra-community links and bridges between communities through the inter-community links. However, bridges are given a higher weight. 
  \begin{equation}
  \small
    \alpha_{Comm}(i) = (1 + \mu_{c_q}) \times \left[ \frac{k_i^{intra}}{max_{(j \in c)}k_j^{intra}} \times R \right] +  (1 - \mu_{c_q})  \times  \left[ \frac{k_i^{inter}}{max_{(j \in c)}k_j^{inter}} \times R \right] ^2
\end{equation}
where $\mu_{c_q} =  \frac{ \sum_{i = 0 \backslash i \in c_q}^{N} k^{inter}_i / k_i}{|c_q|}$  and $R$ is a scaling factor.

\textbf{Community-based Centrality} \cite{zhao2015community}: combines the number of intra-community and inter-community links weighted by the size of their communities.   

   \begin{equation}
       \alpha_{CBC}(i) =  \sum_{c=1}^{N_c} k_{i,c_q} 
\left(
\frac{n_{c_q}}{N}
\right)
   \end{equation}

\textbf{Community‑based Mediator}\cite{tulu2018identifying}: uses the entropy of the intra-community and inter-community links to discover central nodes.
    \begin{equation}
    \small
\alpha_{CBM}(i) = \left[-\sum \rho_i^{intra} log(\rho_i^{intra})] + [- \sum \rho_i^{inter} log(\rho_i^{inter})\right]\ \times \frac{k_i}{\sum_{i=1}^{N} k_i}
\end{equation}
where $\rho_i^{intra} = \frac{k_i^{intra}}{k_i}$ and $\rho_i^{inter} = \frac{k_i^{inter}}{k_i}$.

\textbf{Community Hub-Bridge} \cite{ghalmane2019immunization}: is based on the combination of the number of intra-community links weighted by the size of the community and the inter-community links weighted by the number of neighboring communities.
      \begin{equation}
\alpha_{CHB}(i) = |c_q| \times k_i^{intra} + \sum_{c_l \subset C \backslash  c_q}^{N} \bigvee_{j \in c_l} a_{ij}  \times k_i^{inter}
\end{equation}   

where $\bigvee_{j \in c_l} a_{ij} = 1$ if node $i$ is connected to at least one node $j$ in community $c_l$. 

     \textbf{Modularity Vitality} \cite{modvitality}: is a signed measure that assess the contribution of a node by removing it and measuring the modularity variation. In this study, the absolute value is taken.
\begin{equation}\alpha_{MV}(i) =  M(G_i) - M(G) \end{equation}
where $M$ is the modularity and $M(G_i)$ is the modularity after node $i$'s removal.

   \textbf{Participation Coefficient} \cite{guimera2005functional}: measures the distribution of a node's links across different communities. It's close to 1 if the links are uniformly distributed among all the communities.
  \begin{equation}
\alpha_{PC}(i) = 1 - \sum_{c=1}^{N_c} 
\left(
\frac{k_{i,c_q}}{k_i}
\right)^2
\end{equation}

\textbf{K-shell with Community} \cite{luo2016identifying}: is based on the k-shell hierarchical decomposition node's local and global influence. Both influences are weighted by user-defined parameter $\delta$, set to 0.5 in this study.

\begin{equation}
\alpha_{ks}(i) = \delta \times \beta^{intra}(i) + (1- \delta) \times \beta^{inter}(i) 
\end{equation}
where $\beta^{intra}(i)$ and $\beta^{inter}(i)$ represent the k-shell value of node $i$ by only considering intra-community links and inter-community links, respectively.

\section{Linear Threshold Model}

The Linear Threshold (LT) model describes collective behavior when actors are restricted between two binary alternatives \cite{granovetter1978threshold}. In sociology and economics, it allows modeling, for instance, the decision to buy or not to buy a good. It can also model the diffusion of innovations, rumors, and diseases, strikes, educational attainment, leaving social occasions, voting, migration, and experimental social psychology \cite{granovetter1978threshold}. In all these situations, an actor's decision relies on the fraction of its neighbors choosing one of the two given alternatives. Actors may have different thresholds, which may depend on a multitude of factors. For example, buying a good or service depends on the actor's economic status, age, education level, product utility, etc.

Consider the graph $G=(V, E)$ in which $(u,v) \in E$ is an edge between $u$ and $v$. Every node $v$ possess a threshold $\theta_v \in [0,1]$ and at any time $t$ a node $v$ can be only in one of the two alternative states:
\begin{equation}
v_i(t) = \begin{cases}
      0 & \text{if}\ v_i \ is \ inactive, \\
      1 & \text{if}\ v_i \ is \ active.
    \end{cases}
\end{equation}

Initially, a fraction $X$ of nodes are active. At each time step, an inactive node $v$ check the states of its neighbors. It is activated if the number of active neighbors $m_v$ satisfies:
\begin{equation}
m_v/k_v \geq \theta_v.
\end{equation}

where $k_v$ is the degree of node $v$. The process is iterated until no more node is activated. An active node remains active till the diffusion process ends.

The threshold depicts the node's responsiveness or susceptibility to adopt the alternative chosen by its neighbors. The higher its value, the more restrictive the node becomes in terms of becoming active. One can consider fixed, or random threshold \cite{kempe2003maximizing}. A fixed threshold assumes that individuals adopt a common rule, while a random one allows disparity between the individuals. When there is no a priori information, one commonly uses a uniform distribution.




\section{Datasets and Evaluation Process}
\label{sec:mm}
This section briefly presents the real-world networks under test and the evaluation measures for analyzing and ranking the community-aware centrality measures.

\subsection{Data}
We use 13 real-world networks originating from various domains to investigate the efficiency of the various community-aware centrality algorithms. There are four social networks (Facebook Politician Pages, PGP, Hamsterster, DeezerEU), three biological networks (Human Protein, Interactome Vidal, Kegg Metabolic), three infrastructure networks (U.S. Power Grid, EuroRoad, Internet Autonomous Systems), and three collaboration networks (GrQc, AstroPh, DBLP). Table \ref{TableMacro} summarizes their basic topological properties. Note that as there is no ground truth, Infomap is used to extract their community structure \cite{rosvall2008maps}.

\begin{table}[h]
   \centering
    \caption{Macroscopic topological properties of the real-world networks. \textit{N} is the total number of nodes. $|E|$ is the number of edges. $<k>$ is the average degree.  $<d>$ is the average distance. $\nu$ is the density. $\zeta$ is the transitivity. $k_{nn}(k)$ is the assortativity. * indicates the largest connected component if the network is disconnected.
    }
   \label{TableMacro}     
      \begin{tabular}{lccccccccc}
      \hline 
    Network & $N$ & $|E|$ & $<k>$ &  $<d>$ & $\nu$  & $\zeta$ & $k_{nn}(k)$ \\
    \hline

    EuroRoad* \cite{nr} & 1,039 & 1,305 & 2.51 & 18.39 & 0.002 & 0.035 & 0.090  \\

    Hamsterster* \cite{kunegis2014handbook} & 1,788 & 12,476 & 13.49 & 3.45 & 0.007 & 0.090 & -0.088 \\
    
     Kegg Metabolic* \cite{netz} & 1,865 & 5,769 & 6.19 & 3.12 & 0.003 & 0.030 & -0.224  \\

    Human Protein* \cite{kunegis2014handbook} & 2,217 & 6,418 & 5.78 & 3.84 & 0.002 & 0.007 & -0.331 \\

    Interactome Vidal* \cite{netz} & 2,783 & 6,007 &  4.32 & 4.84 & 0.002 & 0.035 & -0.137 \\

    GrQc \cite{nr} & 4,158 & 13,422 & 6.45 & 6.04 & 0.001 & 0.628 & 0.639 \\
    
    U.S. Power Grid \cite{kunegis2014handbook} & 4,941 & 6,594 & 2.66 & 18.98 & 0.005 & 0.103 & 0.003  \\
    
    Facebook Politician Pages \cite{nr} & 5,908 & 41,729 & 14.12 & 4.66 & 0.002 & 0.301 & 0.018  \\
    
    Internet Aut. Systems \cite{nr}& 6,474 & 12,572 & 3.88 & 3.70 & 0.0006 & 0.009 & -0.181 \\

    PGP \cite{kunegis2014handbook} & 10,680 & 24,316 & 4.55 & 7.48 & 0.0004 & 0.378 & 0.238  \\

    DBLP* \cite{netz} & 12,494 & 49,579 &  7.94 & 4.42 & 0.0006 & 0.062 & -0.046 \\

    AstroPh* \cite{nr} & 17,903 & 196,972 & 22.00 & 4.19 & 0.001 & 0.317 & 0.201  \\

    DeezerEU \cite{rozemberczki2020characteristic} & 28,281 & 92,752 & 6.55 & 6.44 & 0.002 & 0.095 & 0.104  \\

    \hline
      \end{tabular}
\end{table}

\subsection{Evaluation Process}

To perform a comparative analysis of the different community-aware centralities, we perform simulations of the diffusion process on real-world networks. We consider a fixed and random threshold for the Linear Threshold model. We also rank the community network measures using a voting method.





\subsubsection{Activation Size:}In the Linear Threshold model, a node can be one of two states: $(i)$ inactive; $(ii)$ active. At the beginning of the simulation process, the top nodes (fraction $f$ of nodes) ranked according to a given centrality measure are active. At each iteration, activated nodes try to activate their non-active neighbors. The whole dynamical process stops at time $t_f$ when there is no more node activated. $A(t_f)$ represents the final number of activated nodes. It is an indicator of the influence of the initially active fraction at time $t_f$. $A(t)$ increases as $t$ increases, and it remains unchanged after the final time $t_f$. The higher $A(t_f)$, the better the diffusive power of the centrality measure under test in the given network.

\subsubsection{Schulze Voting Method:} Developed in 1997 by Markus Schulze, this ranking method allows selecting a single winner or $k$ top-ranked winners. Given a set S of ballots and C of candidates, every voter gets to express their preference about candidates in the form of a rank-ordered list (ballot). The margin matrix $ M: C \times C $ is constructed using pair of candidates as follows:
\begin{equation}
    M(d,e) = \mathbf{card}(b \ | \ d \succeq_b e ) - \mathbf{card}(b \ | \ e \succeq_b d).
\end{equation} 

Where $d \ and \ e \in C, \mathbf{card}()$ is the cardinality and $\succeq_b$ is the ordering given by ballot $b \in S$. A directed path $p_{(d,e)}$ from candidate $d$ to candidate $e$ is a sequence of candidates $[c_0,...,c_n]$, where $c_0 = d$ and $c_n = e \ (n \geq  0)$. The strength $st$ of path $p_{(d,e)}$ is defined as:
\begin{equation}
         st(p_{(d,e)}) = min\{M(c_i,c_{i+1}) \ | \ 0 \leq i \leq n\}
\end{equation}

A candidate $c \in C$ is said to be winner if and only if: $st(p_{(c,d)}) > st(p_{(d,c)}) $ for every candidate $ d \in C$.
In our setting, the candidates are the community-aware centrality measures, and the voters are the networks. More precisely, each network expresses a vote for each fraction of the initially active nodes. Consequently, the final ranks of the community-aware centrality measures represent a consensus between all the networks for all the values of the fraction of initially active nodes.

\section{Diffusive Power of the Community-aware Centrality Measures}In the Linear Threshold Model, the fixed threshold can vary between $0$ and $1$. The diffusion process unfolds faster and sometimes instantly for a fixed threshold approaching the extreme minimum ($0$). Using a threshold approaching the extreme maximum ($1$) hinders the diffusion process. In this paper, we investigate low, medium, and high threshold values. Due to the lack of space, only results with medium ($\theta_1 = 0.4$) and high threshold ($\theta_2 = 0.7$) are reported. Additionally, we report the results with a random threshold uniformly distributed ($\theta_3 = U[0,1]$).

\begin{figure}[!ht]
\centering
\includegraphics[width=5in, height=5.5 in]{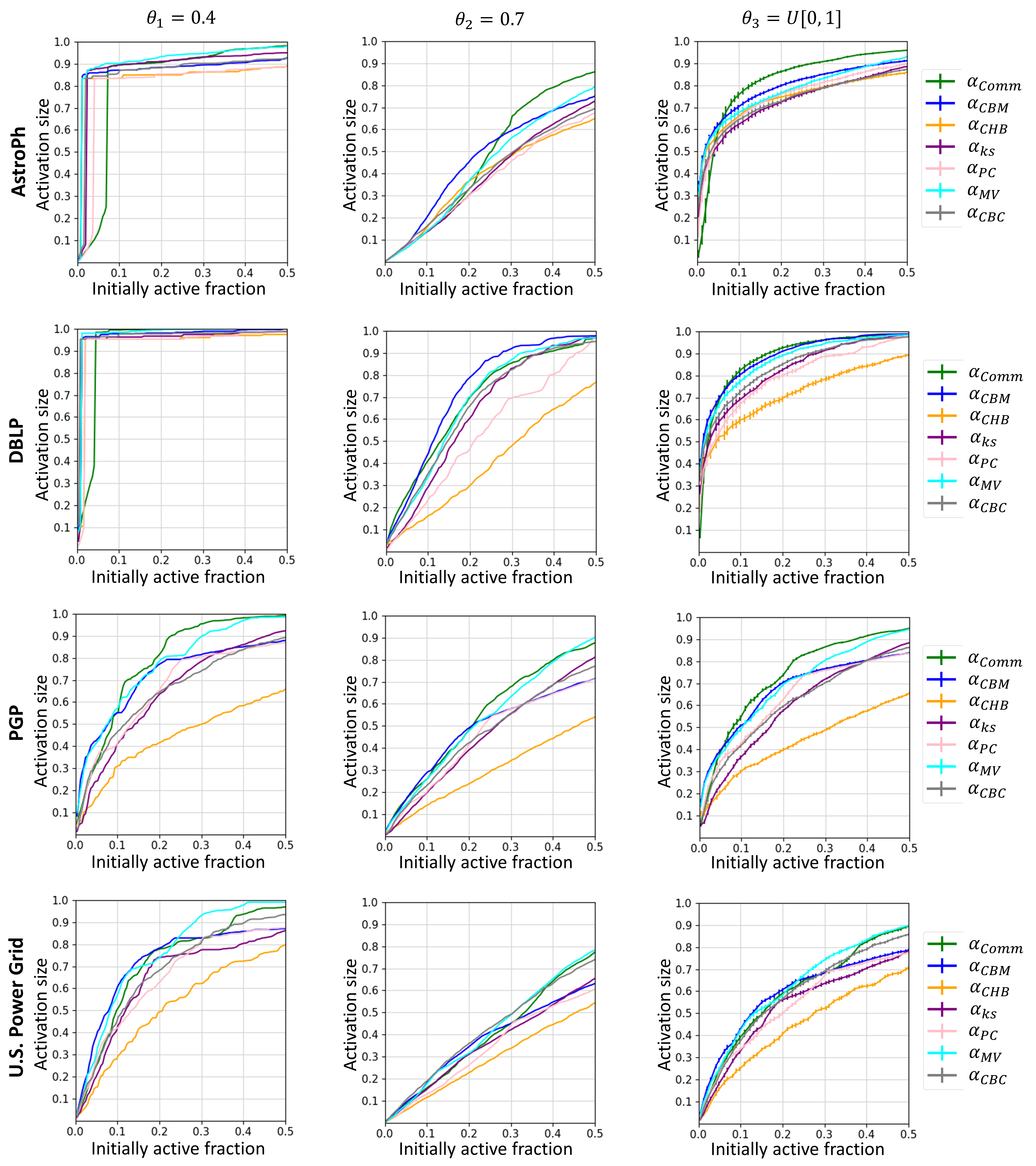}
\caption{The activation size as a function of the fraction of initially active nodes using different thresholds for four real-world networks. The set of the initially active nodes is ordered according to the ranks associated with a given community-aware centrality measure. The deterministic thresholds are $\theta_1 = 0.4$, $\theta_2 = 0.7$, and the random threshold is $\theta_3 = U[0,1]$.}
\label{Fig1}
\end{figure}

\subsection{Comparing the Outbreak Size with a Low Threshold}

A low threshold in the Linear Threshold Model indicates that a node is easily activated. More specifically, a threshold of $\theta_1 = 0.4$ means that if only 40\% of a node's neighbors are activated, the node gets activated. In this situation, one can distinguish two typical behaviors. AstroPh illustrates the first case in Figure \ref{Fig1} on the left first row. One notices an exponential rise in the activation size. Indeed, the activation rate goes from 0\% until almost 90\% when the fraction of initially active nodes is less than 0.1. Afterward, as most of the nodes are activated, there is almost no variation. The networks DBLP, DeezerEU, Hamsterster, Human Protein, Interactome Vidal, Internet AS, and Kegg Metabolic follow the same behavior. It illustrates how effortlessly an epidemic or a rumor can spread in those networks using a very small fraction of initially active nodes. Moreover, suppose we concentrate on the curves when the fraction of initially active nodes is less than 0.1. In that case, it appears that Community-based Mediator and Modularity Vitality slightly outperform the other measures in AstroPh, DBLP, DeezerEU, Hamsterster, and Interactome Vidal.

PGP on the left third row in Figure \ref{Fig1} is a typical example of the second case. Here, a gradual increase in the activation size appears as the fraction of initially active nodes rises from 0\% to 50\%.  The performance of the centrality measures is also well separated at each fraction of initially active nodes. For example, in the PGP network, when the initial proportion of active nodes is 0.3, the maximum activation size reaches almost 95\% when the nodes are active based on the ranks of Comm Centrality. It only reaches 50\% with the initial fraction of Community Hub-Bridge. EuroRoad, Facebook Politician Pages, GrQc, and U.S. Power Grid exhibit similar behavior. The most effective measures in these networks are Community-based Mediator, Comm Centrality, or Modularity Vitality. When the fraction of initially active nodes is low (less than 0.1), Community-based Mediator results in the highest diffusion. In between, the results among the three measures vary. For example, in PGP, Comm Centrality results in the highest activation size from 0.1 till 0.5, and Modularity Vitality joins it starting from a fraction of initially active nodes equaling 0.4.


\subsection{Comparing the Outbreak Size with a High Threshold}


In the Linear Threshold model with a high threshold, nodes are more resistant to change their opinion towards something. A threshold value $\theta_2 = 0.7$ means that the node gets active only if 70\% of its neighbors are active. In this situation, the curves representing the evolution of the activation size as a function of the fraction of initially active nodes are well-separated. One can observe three centralities dominating the other measures for most of the networks under study. They are Community-based Mediator, Comm Centrality, and Modularity Vitality. EuroRoad, Facebook Politician Pages, and U.S. Power Grid depart from this typical behavior. Indeed, Community-based Centrality performs slightly better in these networks.

The effectiveness of the community centrality measures fluctuates with the fraction of initially active nodes. In the low range, Community-based Mediator has the highest activation size. Then as the fraction of initially active nodes grows, either Comm Centrality or Modularity Vitality take the lead. For example, in AstroPh, Community-based Mediator exhibit the highest activation size until the fraction of initially active nodes reaches 0.28. Then, Comm Centrality outperforms it (see the middle first row of Figure \ref{Fig1}). Results are similar for PGP, given in the middle third row of Figure \ref{Fig1}.
Note that only in DBLP and DeezerEU, Comm Centrality results in a slightly higher activation size when the fraction of initially active nodes is less than 0.1. It illustrates that targeting bridges rather than a node with well-mixed intra-community and inter-community links is more effective in these networks.

\subsection{Comparing the Outbreak Size with a Random Threshold Uniformly distributed}
In many cases, the activation threshold of nodes is unknown. Indeed, some people are more or less sensitive to their neighborhood to make a decision. To investigate these kinds of situations, we assign threshold values at random using a uniform distribution. We ran the experiment 50 times with different samples of the threshold sets for each centrality measure. As performance fluctuates between the experiments, we report the average and standard deviation values of the activation size on the right side of Figure \ref{Fig1}.

Results vary with the fraction of initially active nodes. For low fraction values
Community-based Mediator, Comm Centrality, and Modularity Vitality outperform the other measures. More specifically, when the fraction of initially active nodes is lower than 0.2, Community-based Mediator and Modularity Vitality almost overlap. As the fraction of the initially active nodes grows, results show that either Comm Centrality or Modularity Vitality achieve the highest activation size. For example, in AstroPh, above an initial fraction of 0.08, the activation size of Comm Centrality is larger than the other centrality measures. We observe the same behavior for U.S. Power Grid, but it is above an initial fraction value of 0.25, and that is Modularity Vitality that achieves the highest activation size. Comm Centrality joins it at an initial fraction of 0.38.

\section{Ranking the Centrality Measures}
\begin{figure}[!ht]
\centering
\includegraphics[width=5in, height=1.5 in]{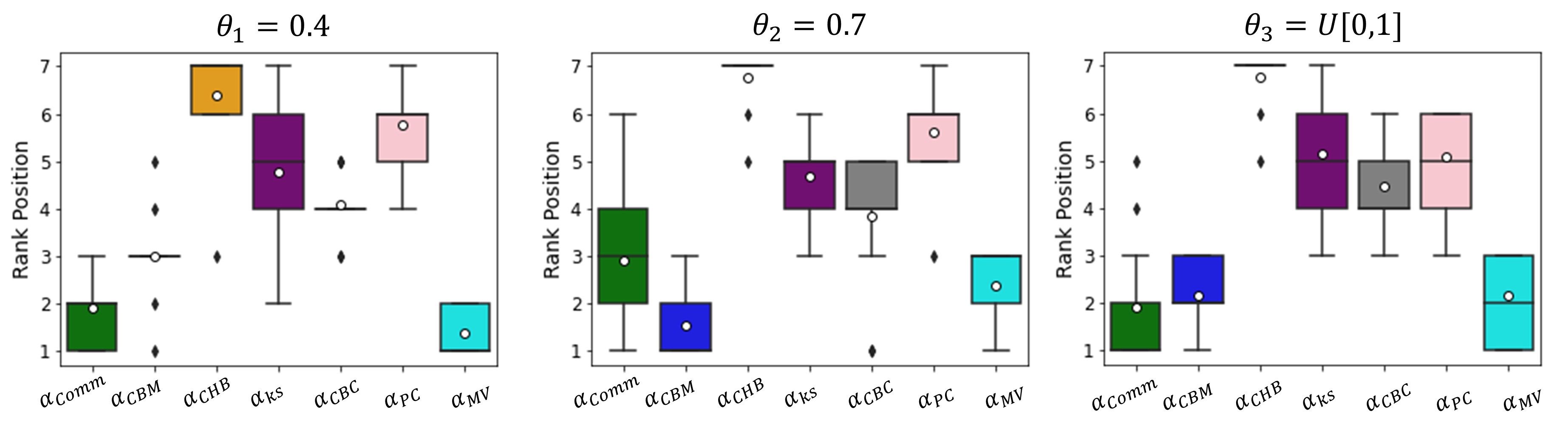}
\caption{The boxplots associated with the ranks of the community-aware centrality measures obtained from the Schulze method. The voting method is computed for each of the thresholds set in the Linear Threshold model ($\theta_1 = 0.4$, $\theta_2 = 0.7$, and $\theta_3 = U[0,1]$). The candidates are the community-aware centrality measures and the voters are the fraction of initially active nodes. The ranks are sorted based on the activation size for each voter.}
\label{Fig2}
\end{figure}

To investigate which centrality measure performs the best across the networks, we utilize the Schulze voting method. The voters are the various fractions of initially active nodes (100 values ranging from $0$ and $0.5$) for each network. We build the preference list (ballot) by sorting the community-aware centrality measures in decreasing order according to the activation size for each voter (initial active fraction). Figure \ref{Fig2} illustrates the results.

When the threshold is low ($\theta_1 = 0.4$), one can see that there is no single winner. Indeed, the first and the second ranks shift between Comm Centrality and Modularity Vitality. Additionally, Community-based Mediator ranks third in most of the networks. Rank four to seven fluctuate between the others. The results show that the means of Comm Centrality and Modularity Vitality fall between the first and second ranks. The average rank of Community-based Mediator and Community-based Centrality lies at the third and fourth rank, respectively. For the other measures, it is between the fifth and seventh ranks. Note that Community-based Mediator and Community-based Centrality exhibit a very low variance while K-shell with Community has the highest. 

When setting the threshold at $\theta_2 = 0.7$, one observes that Community-based Mediator and Community-based Centrality ranks are less stable than a smaller threshold, while the position of Community Hub-Bridge is relatively more stable. According to the mean ranks, Community-based Mediator is between the first and the second, and Comm Centrality ranks between the second and the third. Then Modularity Vitality and Community-based Centrality on the average rank between third and fourth. K-shell with Community, followed by Participation Coefficient and finally Community Hub-Bridge close the rank.  

With a random threshold ($\theta_3 = U[0,1]$), the results don't depart from the ones observed with low and high fixed thresholds. Indeed, similar to the previous results, there's no single winner. The first, second, and third ranks shuffle between Comm Centrality, Community-based Mediator, and Modularity Vitality, while the ranks from fourth till seventh are shared between the rest of the measures.

Overall, Comm Centrality and Modularity Vitality occupy the first two ranks when the threshold is low. In comparison, the position of Community-based Mediator fluctuates between the first two ranks when the threshold is high. The mean rank of the other centralities seems to be relatively insensitive to the threshold variation.

\section{Discussion}

With a deterministic low threshold shared by all the network nodes, one can observe two typical behaviors. The first one is characterized by rapid diffusion to almost all the population with a tiny proportion of active nodes.  In contrast, the activation size rises gradually with the number of initially active nodes in some networks. In these networks, the topology is such that the diffusion is less straightforward. Nevertheless, in both cases, Community-based Mediator, Modularity Vitality, and Comm Centrality outperform the alternative measures. Increasing the threshold limits the spreading. However, the top performers are the same that the ones identified with a low threshold. Finally, when one assigns the threshold at random, the ranks of the top performers vary with the proportion of initially active nodes. When a small fraction of nodes is originally active, Community-based Mediator and Modularity Vitality perform the best. Afterward, Comm Centrality and Modularity Vitality lead to the highest activation size. Hence, the overall behavior of these three community-aware centrality measures is relatively insensitive to threshold variation. The Schulze voting method corroborates these results.

In a previous work, we performed a similar analysis using the SIR propagation model. Results show that Comm Centrality and Community Hub-Bridge perform well in networks with a low fraction of infected/active nodes. In contrast, Modularity Vitality outperforms its alternatives when the initial fraction grows. These results depart from the findings of this paper. Here, when the budget (proportion of initially active nodes) is tight, Community-based Mediator has the highest activation size. When it is higher, Comm Centrality and the absolute of Modularity Vitality perform better.
Moreover, Community Hub-Bridge is one of the worst performers using the LT model. It proves that one must be cautious before extrapolating the findings of such a comparative analysis. Indeed, a well-performing centrality for a given diffusion process is not necessarily as effective with an alternative diffusion model. 

Second, the outperformance of Community-based Mediator when the budget is low suggests that the best strategy is to target nodes with highly unbalanced intra-community and inter-community links. In contrast, targeting bridge-like nodes is more effective with the SIR Model diffusion process.




\section{Conclusion}
\label{sec:conc}
Identifying influential nodes in real-world networks is a major issue. These networks are commonly organized in communities. In contrast with classical centrality measures, recent works incorporate this knowledge to design community-aware centrality measures. Although the propagation process is critical, these works generally rely on the Susceptible-Infected-Recovered model to evaluate the measures. In this work, we investigate the Linear Threshold model as an alternative. We conduct an extensive comparative analysis of seven influential community-aware centrality measures using a set of real-world networks. Results show that Community-based Mediator, Comm Centrality, and Modularity Vitality outperform the alternatives regardless of the threshold. More specifically, Community-based Mediator performs better on a low budget and stubborn nodes, while Comm Centrality and Modularity Vitality show better performance when the budget is higher. The results indicate the importance of investigating measures on various diffusion models to assess the effectiveness and consistency of the centrality measures.
%
%

\bibliographystyle{unsrt}
\bibliography{bibtech}

\end{document}